%%%%%%%%%%%%%%%%%%%%%%%% (plain tex) %%%%%%%%%%%%%%%%%%%%%%%%%%%%%%
\magnification 1200
\font\biga=cmbx10 scaled\magstep3
\font\bigb=cmbx10 scaled\magstep2
\font\sfa=cmr9
\hsize 16.5 true cm
\def\d${ $\displaystyle } 
%%%%%%%%%%%%%%%%%%%%%%%%%%%%%%%%%%%%%%%%%%%%%%%%%%%%%%%%%%%%%%%%%%%%%%%%%%%%%%%
\rightline{\bf DFUB 96-20}
\rightline{\bf TP-USL/96/11}
\rightline{ July 1996}
%%%%%%%%%%%%%%%%%%%%%%%%%%%%%%%%%%%%%%%%%%%%%%%%%%%%%%%%%%%%%%%%%%%%%%%%%%%%%%%
\vskip 1.0 cm

\centerline{\biga BHAGEN-1PH: A Monte Carlo event generator }
\centerline{\biga for radiative Bhabha scattering.}
\vskip 1.0 cm

\centerline{
{\bf M.~Caffo\ $^{ab}$ } and \ 
{\bf H.~Czy{\.z}\ $^c${\footnote 
{ $^{\star}$}{\sfa 
%\baselineskip = .5 true cm \baselineskip = .5 true cm 
     Partly supported by 
     USA-Poland Maria Sk{\l}odowska-Curie
     Joint Fund II (grant MEN/NSF-93-145)
     and
     the Polish Committee for Scientific Research
     (grant no PB659/P03/95/08)
     . } } }
} 

\vskip 1.0 cm
\item{$^a$ \ }{\it INFN, Sezione di Bologna, I-40126 Bologna, Italy }
\item{$^b$ \ }{\it Dipartimento di Fisica, Universit\`a di Bologna, 
             I-40126 Bologna, Italy }
\item{$^c$ \ }{\it Institute of Physics, University of Silesia, 
             PL-40007 Katowice, Poland }

\par
\bigskip

\noindent
{\tt E-mail: \par
caffo@bologna.infn.it \par
czyz@usctoux1.cto.us.edu.pl \par
} 
\bigskip

\vskip 1cm 
\centerline{\bf Abstract }
\par
BHAGEN-1PH is a FORTRAN program providing fast Monte Carlo event generation
of the process $e^+ e^- \mapsto e^+ e^- \gamma$, within electroweak theory,
for both unpolarized beams and also for the longitudinally polarized electron 
beam. 
The program is designed for final leptons outside a small cone around the 
initial leptons direction and has a new algorithm allowing also for a fast 
generation of non collinear initial and final emission, as well as for 
asymmetric and different angular cuts for final leptons.


\vfill \eject
%%%%%%%%%%%%%%%%%%%%%%%%%%%%%%%%%%%%%%%%%%%%%%%%%%%%%%%%%%%%%%%%%%%%%%%%%%%%%%%
\baselineskip = 1 true cm   % double spacing between lines
\centerline{\bigb PROGRAM SUMMARY }
\bigskip
\noindent {\it Title of program:} BHAGEN-1PH

\noindent {\it Program obtainable from:} the authors on request \par\noindent
({\tt E-mail: caffo@bologna.infn.it, czyz@usctoux1.cto.us.edu.pl }).

\noindent {\it Computer/Operating system:} any supporting FORTRAN 77.

\noindent {\it Programming language used:} FORTRAN 77.

\noindent {\it Memory required to execute with typical data:} about 1.6 MB

\noindent {\it No. of bits in a word:} 32

\noindent {\it No. of lines in distributed program:} 3607 lines.

\noindent {\it Subprograms used:} RANLUX [9] (included in the source code). 


\noindent {\it Nature of physical problem:} The process is measured in 
$e^+ e^-$ experiments with $e^-$ beam longitudinally polarized or not,
and is of interest for tests of the Standard Model (QED at low energies), 
as a background estimation for other processes, and as a radiative correction 
contribution to small angle Bhabha scattering for luminosity monitoring. 

\noindent {\it Method of solution:} a Monte Carlo event generation with 
importance sampling method.

\noindent {\it Typical running time:} to generate $10^4$ unweighted events 
are requested 10-20 sec for typical event selection on an AlphaVAX DEC 3000
M700 with OpenVMS 6.2.

\vfill \eject
\baselineskip = .5 true cm   % single spacing between lines

\centerline{\bigb LONG WRITE-UP }
\bigskip
\noindent
1. {\bf Introduction.}
\par
The cross section for $e^+ e^- \mapsto e^+ e^- \gamma$ in QED has been 
calculated with different approximations for various purposes, producing 
distributions to which we refer in [1]. 
At high energies (TRISTAN, LEP, SLC) it is necessary to include the weak-boson 
$Z^0$ exchange and with a longitudinally polarized electron beam (SLC) the 
polarization effects.
For the unpolarized case a relatively simple expression for the square of the 
matrix element is obtained in [2,3], for the longitudinally polarized electron 
beam in [4] is given a reasonably compact expression for the square of the 
matrix element, and in [1] that expression is improved with some other terms 
relevant on the $Z$ boson peak and with all the relevant mass-corrections for 
the configurations in which the final fermions have angles with the initial 
direction larger than, say, 1 mrad.
For extremely forward final fermions the mass-corrections reported there are 
not sufficient. 

Because of the experimental interest in this process we have implemented it 
within a fast Monte Carlo event generator. 
The main target is to obtain a fast generation procedure for the mentioned 
radiative Bhabha scattering configurations, where final leptons are outside 
a small cone around the initial direction (of a semiopening angle of 1 mrad) 
and allowing also for the generation of a non collinear initial or final photon
without loosing the efficiency.
At variance from the existing Monte Carlo event generators for Bhabha 
scattering [5,6,7], 
which include the radiative Bhabha scattering as a correction, 
we keep it as a distinct process and the developed method of generation 
allows to choose asymmetrical and different cuts for electron and positron 
angular variables, notably in t-channel.
  
We do not report here the squared matrix element for the cross-section 
relevant in the region of applicability of the program, which can be found 
in literature for the unpolarized case in [1,3,4] and for the longitudinally 
polarized case in [1].
Here we present the details of the structure of the program BHAGEN-1PH, 
while some results and the numerical comparisons of BHAGEN-1PH with other 
existing programs (for the unpolarized case and special configurations)
are presented in [1]. 
\bigskip

\noindent
2. {\bf Monte Carlo Algorithm.}
\par
The importance sampling method [8] is used as basic Monte Carlo algorithm. 

We denote the differential cross-section as 
$$ d\sigma = \Sigma \quad d\Omega_1 d\Omega_{\gamma} d\omega \ . 
\eqno(2.1) $$
where $d\Omega_1 = d\phi_1 d\cos\theta_1$ 
($d\Omega_{\gamma} = d\phi_{\gamma} d\cos\theta_{\gamma}$) 
is the final electron (photon) solid angle, $\omega$ is the photon energy 
and $\Sigma$ is the distribution.

Moreover we use the following notation: $d\Omega_2 = d\phi_2 d\cos\theta_2$ 
is the final positron solid angle; $E_1 ( E_2)$ is final electron (positron)
energy; $E_b$ is the initial electron and positron energy; $m_e$ is
the electron mass; $s= 4 E_b^2$; $\alpha$ is fine-structure constant.
We work in the CM frame of initial electron and positron with the $z$-axis
chosen along initial electron direction and $x$-$z$-plane given by momenta 
of initial and final electron. 
The last choice is convenient due to the fact that the matrix element
for the unpolarized cross section, as well as the matrix element for
the cross section with a longitudinally polarized beam, depends only on 
one azimuthal angle, while the second one is evenly distributed. 
As a consequence one can actually generate in the chosen frame only four 
variables and the evenly distributed (in the laboratory frame)
azimuthal angle of the final electron can be generated at the end with
the subsequent rotation of all momenta.
  
The distribution $\Sigma$ is approximated by $\Sigma_A$, which consists 
of 10 parts, in the following called channels, related to all possible 
combinations of enhancements due to $Z^0$ resonance, collinear and soft 
photon emission and $t$-channel photon exchange
$$ \Sigma \simeq \Sigma_A = 
   \Sigma_1 + \Sigma_3 + \Sigma_6 + \Sigma_7 + \Sigma_{10} 
   + \left( {{E_1}\over{E_2}} \right)^2
   \left( \Sigma_2 + \Sigma_4 + \Sigma_5 + \Sigma_8 + \Sigma_9 \right) \ . 
\eqno (2.2)$$
With the proper choice of the channel approximants 
$\Sigma_i,\quad i=1,...,10$, described below, 
all the relevant peaks in the distribution $\Sigma$ can be reproduced. 
As each peak can be expressed in a more simple way using different set of 
angular variables, the cross-section (and its channel contributions exact 
$d\sigma_i$ and approximate $d\sigma_i^A$) is rewritten as  
  $$ \eqalign{
  d\sigma = \sum_{i=1}^{10} d\sigma_i & \quad 
          = \sum_{i=1}^{10} {{\Sigma}\over{\Sigma_A}} d\sigma_i^A \cr
          = {{\Sigma}\over{\Sigma_A}} d\omega &\bigl[ 
              \Sigma_1 d\Omega_1 d\Omega_{1\gamma} 
             +\Sigma_2 d\Omega_2 d\Omega_{2\gamma} \cr
   +&\left(\Sigma_3+\Sigma_6+\Sigma_7+\Sigma_{10}\right) 
     d\Omega_1 d\Omega_{\gamma} 
    +\left(\Sigma_4+\Sigma_5+\Sigma_8+\Sigma_9\right)
     d\Omega_2 d\Omega_{\gamma} \bigr] \ , \cr}
\eqno (2.3) $$
where $d\Omega_{1\gamma}$ ($d\Omega_{2\gamma}$) is the photon solid angle
in the frame where $z$-axis is chosen along final electron (positron).
The ratio $w \equiv (\Sigma/\Sigma_A)$, called weight, is a relatively flat 
function. 
This fact and the absorbtion of all the peaks in $\Sigma_1,...,\Sigma_{10}$ 
through an appropriate change of variables, allow for a fast generation. 
Besides of the most relevant peaks, all the other angular and energy 
distributions are absorbed, whenever possible in a simple way.

In the following are given explicit expressions for the approximants 
$\Sigma_i,\quad i=1,...,10$, which define probability density in each channel. 
The integration for each of them is examined and the proper form for 
$d\sigma_i^A$ is obtained. 
The form integrated over one of the azimuthal angles is used, due to its plain 
distribution as previously discussed.

We select channels according to:
\par 1) $s$-channel with final lepton emission (Channels 1-2);
\par 2) $s$-channel with initial lepton emission (Channels 3-8);
\par 3) $t$-channel with photon emission (Channels 9 and 10).

For all the channel approximants we succeeded to obtain analytical integration
(fast generation) and analytically invertible variables.
\bigskip

\noindent 
1) $s$-channel with final lepton emission (Channels 1-2). 

\noindent 
- Channel 1 - final electron emission.
 
To approximate well the peak in the cross-section coming from the $s$-channel 
contribution (from both photon and $Z^0$ boson exchange), when the photon is 
almost collinear to the final electron, we choose 
$$ \eqalign{
  d\sigma_1^A &= \int\limits_0^{2\pi}d\phi_1 \Sigma_1 d\cos\theta_1
    d\Omega_{1\gamma} d\omega \cr
    & = {{\alpha^3}\over{4 \pi s}} \quad F(s) \left( 1+\cos^2\theta_1\right)
        {1\over{a+2\sin^2{{\theta_{1\gamma}}\over{2}}  }} \quad
        {{E_b-{\omega\over 2}}\over{\omega(E_b-\omega)}}d\cos\theta_1
    d\Omega_{1\gamma} d\omega \ , \cr}
\eqno (2.4)$$
where 
$$F(s) = 2 + 4 c_V^2 {{s(s-M_Z^2)}\over{D(s)}} + 2 
  \left[ (c_V^2+c_A^2)^2 -2 P_L c_V c_A (c_V^2+c_A^2) \right]
     {{s^2}\over{D(s)}} \quad ,
\eqno (2.5)$$
$$D(s)=(s-M_Z^2)^2 + M_Z^2 \Gamma_Z^2 \quad ,
\eqno (2.6)$$
$$c_A = {{-1}\over{4 \sin \theta_W \cos \theta_W}} \quad , \quad
  c_V = {{-1+4\sin^2 \theta_W }\over{4 \sin \theta_W \cos \theta_W}} \quad ,
\eqno (2.7)$$
and      
$$ a = {{m_e^2}\over{2 E_b^2}} \quad , 
\eqno (2.8)$$
     
$\theta_W$ is Weinberg angle, $M_Z$ ($\Gamma_Z$) is $Z^0$ boson mass (width)
and $P_L$ is the electron longitudinal polarization.
  
Moreover the expression in (2.4) mimics the electron angular ($\theta_1$) and 
the photon energetic ($\omega$) behaviours of this part of the cross-section. 
The photon energy dependence is well approximated at both ends of its spectrum.

By performing the change of variables
$$x(\omega) = \ln\left( {{\omega}\over{
       \sqrt{E_b^2-\omega E_b} } }\right) 
\quad , \eqno(2.9) $$
$$y(\theta_{1\gamma}) =
     - \ln\left(a+2\sin^2{{\theta_{1\gamma}}\over{2}}\right) 
\quad , \eqno(2.10) $$
$$ z(\theta_1) = \cos\theta_1 + {1\over 3} \cos^3 \theta_1 
\quad , \eqno(2.11) $$
the expression in (2.4) becomes 
$$d\sigma_1^A =
      {{\alpha^3}\over{4 \pi s}} \quad F(s) d\phi_{1\gamma}
      dx(\omega) dy(\theta_{1\gamma}) dz(\theta_1) 
\quad . \eqno(2.12)$$
It is now easy to obtain analytically the integral of (2.12) over the allowed
region of the variables, which is a key step in the importance sampling method 
for the choice of the proper channel for generation.

Also the relations (2.9)-(2.11) are analytically invertible:
$$ \omega(x) = E_b \quad {{2 \exp(x)}
      \over{\exp(x) + \sqrt{\exp(2 x)+4} }}
\quad , \eqno(2.13)$$
$$\sin^2{{\theta_{1\gamma}(y)}\over{2}} = {{\exp(-y) -a}\over 2}
\quad , \eqno(2.14)$$
$$    {\cos\theta_1(z)} = {{3z}\over
  { 1 
 + \left[{ {3\over 2} z 
   +  \sqrt{1+{9\over 4} z^2} }\right]^{2\over3}
 + \left[{ - {3\over 2} z 
   +  \sqrt{1+{9\over 4} z^2} }\right]^{2\over3} 
    } }
\quad . \eqno(2.15)$$
What exposed provides a fast generation in this channel. 

\noindent  
- Channel 2 - final positron emission.
  
This channel approximates well the peak in the cross-section which appears in 
the $s$-channel contribution (from both photon and $Z^0$ boson exchange 
diagrams), when photon is almost parallel to the final positron.  
The situation is analogous to the channel 1, with the electron variables
substituted by the positron ones. 
We use both channels to allow for asymmetrical and different cuts for 
electron and positron.

The formulae are identical to the previous ones (with the change
of subscript 1 into 2) and the approximate expression for channel 2 is 
$$ \eqalign{
  d\sigma_2^A &= \int\limits_0^{2\pi}d\phi_2 \Sigma_2 d\cos\theta_2
    d\Omega_{2\gamma} d\omega \cr
              &= {{\alpha^3}\over{4 \pi s}} \quad F(s) d\phi_{2\gamma}
      dx(\omega) dy(\theta_{2\gamma}) dz(\theta_2) \ . \cr}
\eqno (2.16)$$

The generation is done for convenience in the frame were $x$-$z$-plane is 
given by initial electron and final positron momenta so after generation the 
momenta are rotated into our frame of reference ($x$-$z$-plane given by initial
and final electron).
\bigskip

\noindent  
2) $s$-channel with initial lepton emission (Channels 3-8).
  
The next six channels are used to approximate the initial emission spectrum. 
The situation here is slightly complicated due to the presence
of $Z^0$ boson resonance. 
We choose to use more channels and simpler formulae, instead of a numerical 
approximant, avoiding accuracy adjustments and getting faster procedure.

Channels 5-8 are used only when the total energy is above $Z^0$-boson mass, 
due to the fact that in the photon spectrum there is a peak, connected with 
the $Z^0$-resonance, which is located in the hard part of the spectrum. 
A posteriori is found that the efficiency of the generator improves 
using these channels only when $\tilde \omega = (s-M_Z^2)/(4 E_b) > 1$ GeV.
Covering the phase space with additional four channels allows us to get 
a proper shape of all angular distributions, avoiding the difficulty of 
matching more than one peak with the same variable. 
Furthermore all the variables used are analytically invertible, so the 
generation is fast.  

\noindent  
- Channel 3 - initial electron emission.
  
In this channel the influence of the $Z^0$-boson resonance on the photon
energy distribution is not taken into account. The approximant is therefore 
very similar to the approximants used in channels 1 and 2 and approximates 
well the peak in the cross-section, which appears in the $s$-channel when 
the photon is collinear to the initial electron 
$$ \eqalign{
  d\sigma_3^A &= \int\limits_0^{2\pi}d\phi_1 \Sigma_3 d\cos\theta_1
    d\Omega_{\gamma} d\omega \cr
    & = {{\alpha^3}\over{4 \pi s}} \quad F(s) \left( 1+\cos^2\theta_1\right)
        {1\over{a+2\sin^2{{\theta_{\gamma}}\over{2}}  }} \quad
        {{E_b-{\omega\over 2}}\over{\omega(E_b-\omega)}}d\cos\theta_1
    d\Omega_{\gamma} d\omega \ , \cr}
\eqno (2.17)$$
and the peaks are absorbed in a way analogous to channel 1, with the obvious
substitution $\theta_{1\gamma} \to \theta_{\gamma}$.

\noindent
- Channel 4 - initial positron emission.
 
Similarly to channel 3, the peak is approximated in the $s$-channel 
contribution, when photon is collinear to the initial positron, leaving the 
$Z^0$ resonance as a spectator
$$ \eqalign{
  d\sigma_4^A &= \int\limits_0^{2\pi}d\phi_2 \Sigma_4 d\cos\theta_2
    d\Omega_{\gamma} d\omega \cr
    & = {{\alpha^3}\over{4 \pi s}} \quad F(s) \left( 1+\cos^2\theta_2\right)
        {1\over{a+2\cos^2{{\theta_{\gamma}}\over{2}}  }} \quad
        {{E_b-{\omega\over 2}}\over{\omega(E_b-\omega)}}d\cos\theta_2
    d\Omega_{\gamma} d\omega \ . \cr}
\eqno (2.18)$$
Note however the change in the photon angular dependance, due to the fact 
that the $z$-axis is chosen along initial electron direction.
Now for peak absorption and photon polar angle generation, instead of 
the variable $y(\theta_\gamma)$ in (2.10), we use the variable 
$\tilde y(\theta_\gamma)$ and its inverse 
$$\tilde y(\theta_{\gamma}) = 
\ln\left(a+2\cos^2{{\theta_{\gamma}}\over{2}}\right) \ , \quad 
\cos^2{{\theta_{\gamma}(\tilde y)}\over{2}} = {{\exp(\tilde y) -a}\over 2} \ . 
\eqno(2.19) $$
Generation is done in a frame were $x$-$z$-plane is given 
by initial electron and final positron momenta so again at the end we 
perform the rotation into the proper frame.
  
\noindent  
- Channel 5 - initial electron emission, $Z^0$ resonance and $\theta_2$ 
angular distribution.
  
The influence of the $Z^0$ resonance on the photon energy is accounted for, 
in conjunction with the positron angular distribution and the photon 
emission collinear to the initial electron.
As mentioned this channel is complementary to the channel 3 and of relevance 
for total energy well above the mass of the $Z^0$.
  $$ \eqalign{  
 d\sigma_5^A &= \int\limits_0^{2\pi}d\phi_2 \Sigma_5 d\cos\theta_2
    d\Omega_{\gamma} d\omega \cr
    & = {{\alpha^3}\over{4 \pi s}} \quad c_Z^{-}
        \left( 1+\cos\theta_2\right)^2
        {1\over{a+2\sin^2{ {\theta_{\gamma}} \over{2}}  }} \quad
        {1\over{\tilde \omega}} \quad
        {{16 E_b^2}\over{16(\tilde \omega - \omega)^2 +M_Z^2\Gamma_Z^2}}
        d\cos\theta_2
    d\Omega_{\gamma} d\omega \ , \cr}
\eqno (2.20)$$
where we introduced the notations
$$c_Z^{\pm} = \left(c_V^2+c_A^2\right)^2 \pm 4 c_V^2 c_A^2 \quad , \quad 
\tilde \omega = {{s-M_Z^2}\over{4 E_b}} 
\quad . \eqno (2.21)$$
With the change of variable in (2.10), with the substitution 
$\theta_{1\gamma} \to \theta_{\gamma}$, and the following ones 
$$\tilde z(\theta_2) = {1\over 3}\left(1+\cos\theta_2\right)^3
\ , \eqno (2.22)$$
$$r(\omega) = -{{4E_b^3}\over{M_Z\Gamma_Z\tilde \omega}}\arctan
    \left[{{4(\tilde \omega - \omega)E_b}\over{M_Z \Gamma_Z}}\right]
\ , \eqno(2.23)$$
the following easily integrable expression can be obtained
$$d\sigma_5^A = {{\alpha^3}\over{4 \pi s}} \quad c_Z^{-}
   d\phi_{\gamma} dy(\theta_{\gamma}) d\tilde z(\theta_2) dr(\omega)
\ . \eqno(2.24)$$
These relations have the inverted expressions
$$\cos\theta_2(\tilde z) = {\root 3 \of { 3 \tilde z}} - 1
\ , \eqno (2.25)$$
$${\omega(r) = \tilde \omega - {{M_Z\Gamma_Z}\over{4E_b}}
   \tan\left(-r{{\tilde \omega M_Z\Gamma_Z}\over{4E_b^3}}\right) \ ,}
\eqno (2.26)$$
   allowing for simple generation procedure.

\noindent  
- Channel 6 - initial electron emission, $Z^0$ resonance and $\theta_1$ 
angular distribution.

The influence of the $Z^0$ resonance on the photon energy is accounted 
for this time in conjunction with the electron angular distribution and 
with the photon emission collinear to the initial electron.
Together with channels 3 and 5 gives a complete approximation of the
$s$-channel cross-section with photon emission collinear to the initial 
electron, when the total energy is above the $Z^0$ mass.
  $$ \eqalign{  
 d\sigma_6^A &= \int\limits_0^{2\pi}d\phi_1 \Sigma_6 d\cos\theta_1
    d\Omega_{\gamma} d\omega \cr
    & = {{\alpha^3}\over{4 \pi s}} \quad c_Z^{+}
        \left( 1+\cos\theta_1\right)^2
        {1\over{a+2\sin^2{ {\theta_{\gamma}} \over{2}}  }} \quad
        {1\over{\tilde \omega}} \quad
        {{16 E_b^2}\over{16(\tilde \omega - \omega)^2 +M_Z^2\Gamma_Z^2}}
        d\cos\theta_1
    d\Omega_{\gamma} d\omega \cr
    & = {{\alpha^3}\over{4 \pi s}} \quad c_Z^{+}
      d\phi_{\gamma} dy(\theta_{\gamma}) d\tilde z(\theta_1) dr(\omega)
     \ . \cr}
\eqno (2.27)$$
        
\noindent  
- Channel 7 - initial positron emission, $Z^0$ resonance and $\theta_1$ 
angular distribution.

Now the photon emission collinear to the initial positron is considered, 
accounting for the influence of the $Z^0$ resonance on the photon energy 
and in conjunction with the electron angular distribution, of relevance 
for total energy above the $Z^0$ mass
$$ \eqalign{  
 d\sigma_7^A &= \int\limits_0^{2\pi}d\phi_1 \Sigma_7 d\cos\theta_1
    d\Omega_{\gamma} d\omega \cr
    & = {{\alpha^3}\over{4 \pi s}} \quad c_Z^{-}
        \left( 1-\cos\theta_1\right)^2
        {1\over{a+2\cos^2{ {\theta_{\gamma}} \over{2}}  }} \quad
        {1\over{\tilde \omega}} \quad
        {{16 E_b^2}\over{16(\tilde \omega - \omega)^2 +M_Z^2\Gamma_Z^2}}
        d\cos\theta_1 d\Omega_{\gamma} d\omega \cr
    & = {{\alpha^3}\over{4 \pi s}} \quad c_Z^{-} d\phi_{\gamma} 
      d\tilde y(\theta_{\gamma}) d\bar z(\theta_1) dr(\omega) \ , \cr}
\eqno (2.28)$$
where the variables in (2.19), (2.23) are used with the new variable
 $$\bar z(\theta_1) = -{1\over 3} \left(1-\cos\theta_1\right)^3 \ ,
\eqno (2.29)$$
again invertible into 
$$\cos\theta_1(\bar z) = -{\root 3 \of {- 3 \bar z}} + 1\ . 
\eqno (2.30)$$
        
\noindent  
- Channel 8 - initial positron emission, $Z^0$ resonance and $\theta_2$ 
angular distribution.
 
Again the photon emission is collinear to the initial positron, accounting for 
the influence of the $Z^0$ resonance on the photon energy 
and in conjunction with the positron angular distribution.
This channel with channels 4 and 7 gives a complete approximation
of the $s$-channel cross-section, when the photon is emitted collinear to 
the initial positron and the total energy is above the $Z^0$ mass. 
  
  $$ \eqalign{  
 d\sigma_8^A &= \int\limits_0^{2\pi}d\phi_2 \Sigma_8 d\cos\theta_2
    d\Omega_{\gamma} d\omega \cr
    & = {{\alpha^3}\over{4 \pi s}} \quad c_Z^{+}
        \left( 1-\cos\theta_2\right)^2
        {1\over{a+2\cos^2{ {\theta_{\gamma}} \over{2}}  }} \quad
        {1\over{\tilde \omega}} \quad
        {{16 E_b^2}\over{16(\tilde \omega - \omega)^2 +M_Z^2\Gamma_Z^2}}
        d\cos\theta_2 d\Omega_{\gamma} d\omega \cr
    & = {{\alpha^3}\over{4 \pi s}} \quad c_Z^{+} d\phi_{\gamma} 
        d\tilde y(\theta_{\gamma}) d\bar z(\theta_2) dr(\omega) \ . \cr}
\eqno (2.31)$$
Here the variables in (2.19), (2.23) and (2.29) are used with the 
substitution $\theta_1 \to \theta_2$.   
\bigskip
     
\noindent  
3) $t$-channel with photon emission (Channels 9 and 10).
 
Even if for $s$-channel generation we do not follow any existing solution,
the generation there is relatively simple. The complexity of the
overlapping peaks in $t$-channel makes the problem much more complicated
and the only satisfactory solution to this problem existing up to now [5-7],
is not efficient enough, when one is interested in generating photons outside 
small angular regions around final and initial lepton directions, which is of 
experimental relevance, when the photons are detected separately from leptons. 
We present the method we use for this case in more detail. 
We split the approximant of that part of the cross-section into two parts, 
which are naturally separated in the exact form of the cross-section and 
related to the emission from the electron or the positron line.

\noindent 
- Channel 9 - electron emission. 
   
This part approximates the peaks coming from $t$-channel emission from 
electron line, when photon is mostly collinear to initial or final electron.
$$ \eqalign{  
 d\sigma_9^A &= \int\limits_0^{2\pi}d\phi_2 \Sigma_9 d\cos\theta_2
    d\Omega_{\gamma} d\omega \cr
    & = {{\alpha^3}\over{\pi s}} A
           {{E_b-{\omega\over 2}}\over{\omega(E_b-\omega)}} 
           {1\over{(1+\cos\theta_2)}} \quad
           {1\over{a+1+\cos\theta_{2\gamma}}} \quad
           {1\over{a+1-\cos\theta_{\gamma}}} 
           d\cos\theta_2 d\Omega_{\gamma} d\omega \ , \cr}
\eqno (2.32)$$
  where 
$$ \cos\theta_{2\gamma} = \cos\theta_2 \cos\theta_{\gamma}
     + \sin\theta_2 \sin\theta_{\gamma} \cos\phi_{\gamma} \ .
\eqno (2.33)$$
The factor $A$ will be defined later when its origin will become clear.
\par
With the change of variables (2.9) the approximate expression (2.32) becomes 
$$ d\sigma_9^A  = {{\alpha^3}\over{\pi s}} A \quad
           {1\over{1+\cos\theta_2}} \quad
           {1\over{a+1+\cos\theta_{2\gamma}}} \quad
           {1\over{a+1-\cos\theta_{\gamma}}} 
           d\cos\theta_2 d\Omega_{\gamma} dx(\omega) \ . 
\eqno (2.34)$$
Splitting the angular range of $\phi_\gamma$ into two parts $[0,\pi)$ and
$[\pi,2\pi)$ the variable
$$ v(\phi_\gamma) = {2\over{\sqrt{c^2-b^2}}} 
  \arctan\left[{{\sqrt{c^2-b^2} \tan{{\phi_\gamma}\over 2}}\over{c+b}}\right] 
\ , \eqno (2.35)$$
can be used to absorb one of the peaks.
Here the following notation is used
$$ c = a+1 - \cos\theta_2 \cos\theta_{\gamma} \ , \quad 
b = -\sin\theta_2 \sin\theta_{\gamma} \ . 
\eqno (2.36)$$
The region $[0,\pi)$ is mapped into $v \in [0,v_{max})$,
while $[\pi,2\pi)$ into $v \in [-v_{max},0)$, where 
$$ v_{max} = {\pi\over{\sqrt{c^2-b^2} }} \ . 
\eqno (2.37)$$
All that allows to generate $v \in [-v_{max},v_{max})$ with the inverse 
relations
$$\phi_\gamma = 2 \arctan\left[ {{c+b}\over{\sqrt{c^2-b^2}}}
      \tan\left( {{v\sqrt{c^2-b^2}}\over 2} \right) \right]
      \quad \quad for \quad v \geq 0 
\eqno (2.38)$$
and
$$\phi_\gamma = 2 \pi + 2 \arctan\left[{{c+b}\over{\sqrt{c^2-b^2}}}
      \tan\left({{v\sqrt{c^2-b^2}}\over 2} \right) \right]
      \quad \quad for \quad v < 0  \ . 
\eqno (2.39)$$
To generate in the usual interval the variable $\rho(\phi_{\gamma}) \in (0,1]$ 
is introduced 
$$ \rho(\phi_{\gamma}) = {{v+v_{max}}\over{2v_{max}}} \ , 
\eqno (2.40)$$
so the approximate channel 9 contribution to the cross-section becomes 
$$ \eqalign{  
 d\sigma_9^A  &= {{2\alpha^3}\over{s}} A \quad
           {1\over{1+\cos\theta_2}} \quad
           {1\over{a+1-\cos\theta_{\gamma}}} \cr
 & {{d\cos\theta_2 d\cos\theta_{\gamma} d\rho(\phi_{\gamma}) dx(\omega)}
   \over{\sqrt{a^2+2a(1+\cos\theta_2 \cos\theta_{\gamma})
   +(\cos\theta_2 + \cos\theta_{\gamma})^2}  }} \ . \cr }
\eqno (2.41)$$
To absorb the remaining peaks and to eliminate the square root in the 
denominator the following change of variables is done 
$$ z_1(\theta_{\gamma}) = a+1-\cos\theta_{\gamma} \ , 
\eqno (2.42)$$
and
$$z_2(\theta_2) = 1+\cos\theta_2 \ , 
\eqno (2.43)$$
so to have
$$ \eqalign{  
d\sigma_9^A  &= - {{2\alpha^3}\over{s}} A \quad
           {1\over{z_2(\theta_2)}} \quad
           {1\over{z_1(\theta_{\gamma})}} \cr
& {{dz_2(\theta_2) dz_1(\theta_{\gamma}) d\rho(\phi_{\gamma}) dx(\omega)}
\over{\sqrt{z_1^2(\theta_{\gamma})+z_2^2(\theta_2)
-2(a+1)z_2(\theta_2) z_1(\theta_{\gamma}) 
-2 z_2(\theta_2) +2 (a+1)^2z_2(\theta_2) }  }}  \ . \cr }
\eqno (2.44)$$
With the further change of variables
$$ t(\theta_{\gamma},\theta_2) = {{D+z_1(\theta_{\gamma})
+\sqrt{\left(D+z_1(\theta_{\gamma})\right)^2+4P}}\over{2\sqrt{P}}} \ , 
\eqno (2.45)$$
whose inverse is      
$$z_1(\theta_{\gamma}) = \sqrt{P}\left(t-{1\over t}\right) - D \ , 
\eqno (2.46)$$
with
$$ D = -(a+1) z_2(\theta_2) \ , \quad 
   P = {1\over 4} a (2+a) z_2(\theta_2) (2-z_2(\theta_2)) \ , 
\eqno (2.47)$$
the approximate expression becomes      
$$ d\sigma_9^A  = - {{2\alpha^3}\over{s}} A \quad
            {1\over{z_2(\theta_2)}} \quad     
            {1\over{\sqrt{P}t^2 -Dt-\sqrt{P} }}\quad
dz_2(\theta_2) dt(\theta_{\gamma},\theta_2) d\rho(\phi_{\gamma}) dx(\omega)
\ . \eqno(2.48)$$
The last change of variables is 
$$r_1(\theta_{\gamma},\theta_2) = {{r-r_{min}}\over{r_{max} - r_{min}}} \ , 
\eqno (2.49)$$
and     
$$r_2(\theta_2) = {1\over {z_2(\theta_2)}} \ , 
\eqno (2.50)$$
where
$$r = {1\over{\sqrt{\Delta}}}
\ln{\biggl\vert {{2\sqrt{P} t -D - \sqrt{\Delta}}\over
{2\sqrt{P} t -D + \sqrt{\Delta}}} \biggr\vert} \ , 
\eqno (2.51)$$
and            
$$ \Delta = z_2^2(\theta_2) + 2 a (2+a) z_2(\theta_2) \ , 
\eqno (2.52)$$
so that the minimum and maximum values of $r_{m}$ ($m$ stands for $min$ 
or $max$) are 
$$ r_{m} = {1\over{\sqrt{\Delta}}}      
\ln{\Biggl\vert {{2 z_{1m} \left[z_{1m} - (a+1) z_2(\theta_2) +
\sqrt{z_{1m}^2 -2z_{1m}(a+1)z_2(\theta_2) +\Delta} \right]} \over
{\left( z_{1m}+\sqrt{\Delta} 
+ \sqrt{z_{1m}^2 -2z_{1m}(a+1)z_2(\theta_2) +\Delta} \right)^2}} \Biggr\vert}
\ , \eqno (2.53)$$
and the approximate expression is now          
$$ d\sigma_9^A  =  {{2\alpha^3}\over{s}} A \quad (r_{max} - r_{min}) z_2
dr_2(\theta_2) dr_1(\theta_{\gamma},\theta_2) d\rho(\phi_{\gamma}) dx(\omega)
\ . \eqno(2.54)$$
All the variables used are analytically invertible 
$$ \eqalign{  
       \cos\theta_2 &= -1 +{1\over{r_2}} \ , \cr
                  r &= r_{min} + r_1 (r_{max} - r_{min}) \ , \cr
\cos\theta_{\gamma} &= a+1-{{2\Delta e^{r\sqrt{\Delta}}}\over
        { \left(1 - e^{r\sqrt{\Delta}}\right)\left[
       \sqrt{\Delta}\left(e^{r\sqrt{\Delta}}+1\right)
            + D \left(1-e^{r\sqrt{\Delta}}\right) \right]}} \ . \cr}
\eqno (2.55)$$
A convenient choice for the factor $A$ is then 
$$ A = {400\over{(r_{max} - r_{min}) z_2(\theta_2)}} \ , 
\eqno (2.56)$$
so to have at last the very simple approximate expression
$$ d\sigma_9^A  =  {{800\alpha^3}\over{s}} \quad
        dr_2(\theta_2) dr_1(\theta_{\gamma},\theta_2)
        d\rho(\phi_{\gamma}) dx(\omega) \ . 
\eqno(2.57)$$
The choice for $A\ne 1$ in (2.56) implies that the approximation of the peaks 
in $t$-channel is not perfect, but nevertheless is good enough (logarithmic in 
$z_2(\theta_2)$) to allow a fast generation. 
Moreover it is chosen to fix properly the relative strength respect to the 
other channels subgenerators.
    
\noindent 
- Channel 10 - positron emission. 
   
This part is specular to the channel 9 and approximates the peaks
coming from $t$-channel emission of a photon mostly collinear to initial 
or final positron
$$ \eqalign{  
 d\sigma_{10}^A &= \int\limits_0^{2\pi}d\phi_1 \Sigma_{10} d\cos\theta_1
    d\Omega_{\gamma} d\omega \cr
    & = {{\alpha^3}\over{\pi s}} {\tilde A}
           {{E_b-{\omega\over 2}}\over{\omega(E_b-\omega)}} 
           {1\over{1-\cos\theta_1}} \quad
           {1\over{a+1+\cos\theta_{1\gamma}}} \quad
           {1\over{a+1+\cos\theta_{\gamma}}} 
           d\cos\theta_1 d\Omega_{\gamma} d\omega  \cr
   & =  {{800\alpha^3}\over{s}} \quad
           dr_2(\theta_1) dr_1(\theta_{\gamma},\theta_1)
        d\rho(\phi_{\gamma}) dx(\omega)
     \ , \cr}
\eqno (2.58)$$
  where now
$$z_2(\theta_1) = 1-\cos\theta_1 \ , \eqno (2.59)$$
and   
$$ z_1(\theta_{\gamma}) = a+1+\cos\theta_{\gamma} \ , \eqno (2.60)$$
with
$$c=a+1 - \cos\theta_1 \cos\theta_{\gamma} \ , \eqno (2.61)$$
and      
$$ b = -\sin\theta_1 \sin\theta_{\gamma} \ , \eqno (2.62)$$
while all the other expressions remain valid. 
$\tilde A$ has the same functional form of $A$, but as $z_2$, $z_1$, $c$ and
$b$ are now different, it is different from $A$.
\bigskip

Having so given the separate description of various channels, obtaining 
a good approximant of the cross-section, the events can be generated in an 
easy way according to the approximated forms in each channel.
  
The pseudo random number generator used to produce the equidistributed random 
numbers between [0,1) is the CERNLIB generator (V115) RANLUX [9], which source 
code is added into our program.
  
The algorithm used to select the channel for generation is the following:
the $i$-th channel is chosen randomly with the probability
$$ P_i = {{v_i}\over{\sum_{j=1}^{10}{v_j}}} \ , \eqno (2.63)$$
 where
$$v_i = \int_{{\bf V}_i} d\sigma_i^A \ , \eqno (2.64)$$
and ${\sl V}_i $ is the allowed region of the variables appearing in
$d\sigma_i^A$.
  
A random generation of 1000 events is done at the beginning and the 
maximum-weight $w_S$ is taken to be 2.5 times the largest 
of the generated weights $ w = {{\sigma}\over{\sigma_A}}$. 
As the $w$ function is relatively flat the method works well and in all 
tests an event with the weight $w > w_S$ was never generated. 
Nevertheless this eventuality is checked automatically in the program and 
in case events with $w > w_S$ appear the cross-section corresponding to them 
is given in the output. However the cross-section and its error calculated 
with the weighted event sample is not affected by the appearance of the events
with weight $w > w_S$.

The cross-section $\sigma$ integrated over the allowed phase space 
is calculated as a sum of all channels contributions, according to the 
stratification method described in [8], 
$$ \sigma = \sum_{i=1}^{10} \sigma_i = \sum_{i=1}^{10} \int d\sigma_i 
          = \sum_{i=1}^{10} \int {{\Sigma}\over{\Sigma_A}} d\sigma_i^A \ , 
\eqno (2.65)$$
and its statistical error $\Delta \sigma$ is calculated as
$$ \Delta \sigma = \sqrt{\sum_{i=1}^{10} (\Delta\sigma_i )^2 } \ , 
\eqno (2.66)$$
where $\Delta \sigma_i$ is the statistical error of the $i$-th channel
contribution $\sigma_i$.
  
A more complicated event selection, then that allowed by choosing the range 
of the generation variables, can be implemented by simply setting the 
weight of the generated event to zero in case it is not allowed by the 
requested cuts (this can be easily done by the user in the subroutine TRIGGER).

Finally we remark that a big effort was devoted to assure that the accuracy 
of the calculations is not lost due to possible cancellations in formulae.
\bigskip

\noindent
3. {\bf Routines in the Generator.}
\par

\noindent BHAGEN-1PH \par
It is the main program and calls the subroutines PARAMET, VOLUME, MAXEST, 
GENER and CROSS.
It calculates and writes the total cross section for both weighted
and unweighted events sample.

\noindent PARAMET(i1,i2,i3) \par
The argument i1 is the logical number of the input file, from which all 
relevant parameters are read (see input file description in the Users Guide 
section); i2 is the logical number of the first output file 
(BHAGEN-1PH.RES), where parameters and cross-section values are reported; 
i3 is the logical number of the second output file (BHAGEN-1PH.EVN), 
where generated events are listed (if requested).
The values, which are constant during all the run, are calculated 
(such as coupling constants, boundaries of the integration regions etc.) 
and the random number generator is initialized.

\noindent VOLUME \par

The values of $v_i$ and $P_i$ ($i=1, ... , 10$) are calculated 
(see algorithm description).

\noindent MAXEST \par

It generates 1000 events in the same way as in GENER and on its basis 
estimates the maximal weight $w_S$.

\noindent GENER \par

It chooses randomly (RANDOMN1) with the probability $P_i$ a given channel
and calls a proper generation procedure in each channel (GENER1, ..., GENER10). 
It writes the generated events (kinematics of the outgoing particles), if 
requested. It stops the generation, when the unweighted event sample is equal 
to the requested number.
   
\noindent CROSS \par

It calculates the integrated cross-sections for each channel, for both 
weighted and unweighted event sample.

\noindent GENER1, ..., GENER10 \par

They are master programs for each separate channel. They generate (RANDOMN)
the kinematical variables, which are chosen as independent (see algorithm
description). The subroutines GENER3, ..., GENER10 call the subroutines 
PHASESP1 or PHASESP2 to calculate the rest of the necessary kinematics 
(energies and angles of the final particles) and reject the event if it is 
generated outside the allowed phase space. 
In GENER1 and GENER2 the same is done inside the subroutines, as the 
calculations are different from other channels. 
All the subroutines call FRAME, which generate (RANDOMN1) the azimuthal angle 
of the final electron and rotates the event (up to this call the event is 
given in a frame where the $x$-$z$-plane is given by initial and final 
electron momenta). 
They also call TRIGGER, which is the event selection subroutine to be supplied 
by the user, presently with no action.
All the subroutines call FRATIO3, where the weights are calculated and summed 
and where the hit or miss method is applied, selecting unweighted event sample.

\noindent RANDOMN1 \par
  
It gives one random number evenly distributed on (0,1] used to choose the 
channel or to generate the azimuthal angle of the final electron. 

\noindent RANDOMN \par

It gives a sequence of 5 random numbers evenly distributed on (0,1]: 4 to 
generate the event variables and 1 for the selection of the unweighted event.  

\noindent PHASESP1 and PHASESP2 \par

They calculates energies and angles of the final particles, given
$\phi_1, \theta_1, \phi_{\gamma}, \theta_{\gamma}$ and $\omega$ (PHASESP1), 
or $\phi_2, \theta_2, \phi_{\gamma}, \theta_{\gamma}$ and $\omega$ (PHASESP2).

\noindent FRAME \par

The azimuthal angle of the electron is generated randomly (RANDOMN1) and the 
event (momenta of the final particles) is rotated around $z$-axis by this angle.
This allows the use of an arbitrary frame with $z$-axis along the initial
electron. Before that rotation the $x$-$z$-plane is the plain given by 
initial and final electron momenta.

\noindent TRIGGER \par

This is an event selection subroutine, to be supplied by the user, allowing 
for rejection (k=0 has to be set inside the subroutine) or acceptance 
(k=1 has to be set inside the subroutine) of the generated event.
The final particles energies and angles are stored in the "ev" vector.
They are given in CM frame of initial electron and positron, with
the $z$-axis along initial electron momentum.
If left as it is given in the distributed version, the trigger has no effect 
on the generation.
The reading of the event vector is as follow: 
 
   ev(1)  - the electron energy in GeV

   ev(2)  - the electron polar angle in rad.

   ev(3)  - the electron azimuthal angle in rad.

   ev(4)  - the positron energy in GeV

   ev(5)  - the positron polar angle in rad.

   ev(6)  - the positron azimuthal angle in rad.

   ev(7)  - the photon energy in GeV

   ev(8)  - the photon polar angle in rad.

   ev(9)  - the azimuthal angle in rad.

\noindent FRATIO3 \par

It calculates parameters which enter in both, the exact and the
approximate cross-sections. 
It calculates the weight ($w$) of the event and it provides the book keeping 
for the calculation of the cross-section for the weighted sample, adding $w$ 
to $\sum_{i} w_i$ and $w^2$ into $\sum_{i} w^2_i$ for the proper channel.
It checks if $w$ is smaller of the estimated maximum-weight ($w_S$), and if not,
it sums the weight $w \equiv w^o_i$ into the separate sums $\sum_{i} w^o_i$ and 
$\sum_{i} (w^o_i)^2$ allowing for the calculation of a separate cross-section 
of such overweighted events, writing also identifying parameters. 
The unweighted events sample can still be used, if the contribution to the 
cross-section due to the overweighted events is within the errors of the total 
cross-section. However it never happened to find an event with weight $w > w_S$
in the test runs, so the maximum estimation is pretty good. 
The hit or miss method is also applied  here: the randomn number $r5$ is 
generated (RANDOMN), $w_t = r5 \ w_S$ is calculated and the event is 
accepted if $w > w_t$. 

\noindent RFULL \par

It calculates the exact differential cross-section; some factors which are 
the same in both exact and approximate form are omitted, 
as only the ratio of the two is used in calculations.

\noindent CHANNEL1, ..., CHANNEL10 \par

They calculate the approximate differential cross-sections for each channel
(but the factors which are the same in both exact and approximate form, 
as only the ratio of the two is used in calculations). 

\noindent FACT \par

It is an auxiliary function used by CHANNEL9 and CHANNEL10.
\bigskip
  
\noindent
4. {\bf Users Guide.}
\par

In the version installed on VAX with VMS system the parameters
are given in the BHAGEN-1PH.COM file (the same file where are the system
commands running the program). 
In the version installed on IBM 340 (RISC/6000) with the system AIX 3.2, 
the only differences can be in the syntax of the OPEN statement (depending 
on the FORTRAN 77 version installed) and in the fact 
that the file BHAGEN-1PH.COM contains only parameters. 
The user has to supply the following data:

\noindent - The mass of the $Z^0$ boson and its width.

\noindent  - The mass of the $W$ boson and the value of $\sin^2\theta_W$. 
Actually only one of this parameters is used by the program. 
If $\sin^2\theta_W$ is set to zero the program uses the expression 
$\sin^2\theta_W = 1 - (M_W^2/M_Z^2)$ to calculate it, while
if it is given different from zero, $M_W$ is not used by the program.
Up to now no radiative corrections are included into the program.

\noindent - The beam energy.

\noindent - The degree of longitudinal polarization of the electron beam. 

\noindent - The number of requested events in the unweighted events sample.

\noindent - The minimal and maximal allowed energy of the final particles.
The maximal allowed energy of the photon can be modified by the program 
(and notified in the output file) to
$E_{\gamma}^{max} = 2 E_b - E_1^{min} - E_2^{min} \le E_b (1 - m_e^2/E_b^2)$
if the given maximum is too high to be consistent with the other cuts supplied 
by the user.

\noindent - The minimal and maximal polar angles (in radians) of the final 
electron $(\theta_- \equiv \theta_1)$ and 
positron $(\theta_+ \equiv \pi - \theta_2)$ with their initial directions.

\noindent - The minimal angle between final lepton and photon 
(it can be set to zero).

\noindent - The minimal angle between initial lepton and final photon
(it can be set to zero).

\noindent - The maximal allowed acollinearity (acollinearity = 
$ \mid \pi - \theta_1 -\theta_2 \mid $, where $\theta_1$ and $\theta_2$ 
are the angles between final electron and positron momenta with the initial
direction of the electron) and acoplanarity (acoplanarity = 
$\mid \pi - (\phi_2 -\phi_1)\mid$), $\phi_1$ and $\phi_2$ are the azimuthal 
angles of the final electron and positron.

\noindent - A flag which allows writing (1) or no writing (0) the generated
events into the created file BHAGEN-1PH.EVN.

\noindent - A flag which allows writing (1) or no writing (0) the separate
channel contributions.

\noindent - A flag which allows to take into account separate contributions 
from $s$- and $t$-channel, which can be useful for tests, but require 
the switch to be set to 0, for comparing the results with experiment. 
If the flag is equal to 0 full version of the program is used; if it is equal
to 1 only $t$-channel photon exchange contribution is included (with up-down 
interference neglected); if it is equal to 2 only $s$-channel contribution 
is included (from both $\gamma$ and $Z^0$ exchange diagrams).

\noindent - A seed to initialize the pseudo random number generator.

Two output files are created. 
The first one BHAGEN-1PH.RES contains the calculated cross-sections
for both weighted and unweighted event sample and if requested
the contributions from all ten channels.
The second one BHAGEN-1PH.EVN contains the generated events (if the 
proper flag was set to 1). 
Both files report the parameters, which are read from the input file.
\bigskip
%\vfill\eject

\noindent{\bf Acknowledgments}\par
Useful discussions with E. Remiddi are gratefully acknowledged. 
One of us (HC) is grateful to the Bologna Section of INFN and to the Department 
of Physics of Bologna University for support and kind hospitality.
\vfill \eject

\baselineskip = 1 true cm   % double spacing between lines
\noindent {\bigb References.} 
\bigskip
\item{[1]\ } M. Caffo, H.Czy{\.z} and E. Remiddi, Phys. Lett. B 378 (1996) 357.

\item{[2]\ } F.A. Berends et al., Phys. Lett. B 103 (1981) 124.

\item{[3]\ } F.A. Berends et al., Nucl. Phys. B206 (1982) 61.

\item{[4]\ } M. Caffo, R. Gatto and E. Remiddi, Nucl. Phys. B286 (1987) 293. 

\item{[5]\ } F.A. Berends, R. Kleiss, Nucl. Phys. B228 (1983) 537. 

\item{[6]\ } S. Jadach et al., Phys. Lett. B 253 (1991) 469.

\item{[7]\ } F.A. Berends, R. Kleiss and W. Hollik, Nucl. Phys. B304 (1988) 712.

\item{[8]\ } J.M. Hammersley and D.C. Handscomb, {\it Monte Carlo Methods},
             London:Methuen (1964), \par
             F. James, Rep. Prog. Phys. 43 (1980) 1145.
\item{[9]\ } F. James, Comp. Phys. Comm. 79 (1994) 111.

\vfill \eject

\baselineskip = .5 true cm   % single spacing between lines
\noindent {\bigb TEST RUN INPUT.}
\bigskip

\noindent {\bf Input file to run with VMS system:}
\bigskip
\parindent=0mm
{\obeylines {\obeyspaces {\tt 

\$run bhagen-1ph.exe
91.1887             Z mass  (GeV)
2.49661             Z width (GeV)
80.304              W mass (GeV) (used only if sin**2(thW)=0.) 
0.224482            sin**2(thW)(if=0 then it is calculated =1-(M\_W/M\_Z)**2
47.5                e\_beam (GeV)
1.0                 longitudinal polarization of the initial electron
10000               nevent - numb. of events in unweighted event sample
4.75   23.75        eg\_min,eg\_max (GeV)      see comment (1)
23.75   47.5        e1\_min,e1\_max (GeV)      see comment (2)
23.75   47.5        e2\_min,e2\_max (GeV)      see comment (3)
0.7   2.3           th1\_min,th1\_max (rad)    see comment (4)
0.8   2.2           th2\_min,th2\_max (rad)    see comment (5)
0.01                th1g\_min=th2g\_min (rad)  see comment (6)
0.1                 thg\_min (rad)             see comment (7)
180. 180.           acolcut(deg),acoplcut(deg) see comment (8)
1                   printing(1) no printing(0) generated events
0     printing(1)/no-printing(0) info on separete channel contrib.
0       s-t channel switch see comment (9)
123456789  seed to the random number generator
\$exit

 Comments:

(1) - min. and max. photon energy in GeV. The minimum photon energy
      cannot be set to 0 !!!! The maximum photon energy is set to
      eg\_max =  e\_beam * (1 - me**2/e\_beam**2) in case it is given bigger
      then that value. This is the biggest value of the photon energy
      allowed by four momentum conservation.
(2) - min. and max. final electron energy in GeV. In case e1\_min=0.
      it is set to the electron mass by the program.
(3) - min. and max. final positron energy in GeV. In case e2\_min=0.
      it is set to the electron mass by the program.
(4) - min. and max. electron polar angle measured from initial 
      electron direction. 
      A user is kindly requested not to use this program with 
      th1\_min < 0.001 rad. as the formulae used are not adequate
      for that region.
(5) - min. and max. positron polar angle measured from initial 
      positron direction.
      A user is kindly requested not to use this program with 
      th2\_min < 0.001 rad. as the formulae used are not adequate
      for that region.
(6) - minimal angle between final electron(positron) and the photon.
      In case they are not equal the smaller should be given.
      It can be set to 0.
(7) - minimal angle between initial electron(positron) and the photon.
      In case they are not equal the smaller should be given.
      It can be set to 0.
(8) - maximal acolinearity and acoplanarity allowed.
      Where acolinearity is defined as |(pi - th1 - th2)|, and th1(th2) 
      is the angle between final electron(positron) and initial electron.
      While acoplanarity is defined as |(phi2-phi1-pi)|, where
      phi1 and phi2 are respectively the electron and positron 
      azimuthal angles.
(9) - 0:all included;
      1:t-channel QED only with up-down interference neglected
        suitable for tests at small angles
      2:s-channel only, both photon and Z boson exchange include
----------------------------------------------------------------------
 Additional, more detailed, event selection can be done by modification 
 of the subroutine TRIGGER in the FORTRAN code.  
 The cross section given in the output corresponds to the complete event
 selection e.g. with the cuts imposed by both the input parameters 
 in this file and the cuts imposed in the subroutine TRIGGER.

 A simple example how to use this subroutine is given below.

 An event with electron having energy between 0.1 GeV and 10 GeV is
 rejected and accepted if the electron energy is outside this region.

c * * * * * * * * * * * * * * * * * * * * * * * * * * * * * * * * * *
c
c   This is a trigger which is to be defined by user allowing for the
c  rejection (k=0 has to be set) or acceptance (k=1 has to be set) 
c  of the generated event.
c   The final particles energies and angles are stored in the 'ev' vector.
c  They are given in CM frame of initial electron and positron, with
c  the z-axis along initial electron momentum.
c   
c 
c   ev(1)  - electron energy in GeV
c   ev(2)  - theta (electron) in rad.
c   ev(3)  - phi (electron) in rad.
c   ev(4)  - positron energy in GeV
c   ev(5)  - theta (positron) in rad.
c   ev(6)  - phi (positron) in rad.
c   ev(7)  - photon energy in GeV
c   ev(8)  - theta (photon) in rad.
c   ev(9)  - phi (photon) in rad.
c
      subroutine trigger(ev,k)
c
      implicit real*8 (a-h,o-z)
      dimension ev(9)
c
      if((ev(1).gt.0.1d0).and.(ev(1).lt.10.d0))then
        k=0
      else
        k=1
      endif
c
      return
      end
c * * * * * * * * * * * * * * * * * * * * * * * * * * * * * * * * * *
}}}
\vfill \eject
\noindent {\bigb TEST RUN OUTPUT.}
\bigskip

\noindent {\bf Output file BHAGNE-1PH.RES.}
\bigskip

{\obeylines {\obeyspaces {\tt 

 BHAGEN-1PH,   VERSION OF 22-JAN-1996                        

    Z mass (GeV) :   91.188700D+00
    W mass (GeV) :   80.304000D+00
  sin**2 (th\_w) =   0.224482000000000     
  Total Z width (GeV) =   2.4966100D+00
  E\_beam  =  47.5000 GeV
 Longitudinal polarization of electron =   1.0000
   23.7500D+00 GeV <= electron energy <=   47.5000D+00 GeV
   23.7500D+00 GeV <= positron energy <=   47.5000D+00 GeV
  4.75000D+00  GeV <  photon energy  <   2.37500D+01  GeV 
  7.00000D-01  rad  <  theta (electron)  <   2.30000D+00  rad
  9.41593D-01  rad  <  theta (positron)  <   2.34159D+00  rad
  Minimal allowed angle between photon and final lepton  =  1.00000D-02 rad.
  Minimal allowed angle between photon and initial lepton=  1.00000D-01 rad.
  acollinearity <   3.14159D+00 rad.
  acoplanarity  <   3.14159D+00 rad.

 * * * * * * * * * * * * * * * * * * * * * * *

 no. of events with weights > max.weight:      0

Total cross section of the events with weights > max. weight :

 cross section (nb) =   0.0000000D+00  +-  0.00D+00

 * * * * * * * * * * * * * * * * * * * * * * *

 Number of generated (unweighted) events in the sample:    10000

 Number of hits (weighted events) in the sample:          130202

 Summary : 

Total cross section obtained using weighted events :

 total cross section1 (nb) =   1.3745087D-02  +-  3.61D-05

Total cross section obtained using unweighted events:

 total cross section2 (nb) =   1.3874919D-02  +-  1.39D-04

}}}
\vfill \eject
\noindent {\bf Output file BHAGNE-1PH.EVN - only first two events are reported.}
\bigskip

{\obeylines {\obeyspaces {\tt 

 BHAGEN-1PH,   VERSION OF 22-JAN-1996                        

    Z mass (GeV) :   91.188700D+00
    W mass (GeV) :   803.04000D-01
  sin**2 (th\_w) =   0.224482000000000     
  Total Z width (GeV) =   2.4966100D+00
  E\_beam  =  47.5000 GeV
 Longitudinal polarization of electron =   1.0000
   23.7500D+00 GeV <= electron energy <=   47.5000D+00 GeV
   23.7500D+00 GeV <= positron energy <=   47.5000D+00 GeV
  4.75000D+00  GeV <  photon energy  <   2.37500D+01  GeV 
  7.00000D-01  rad  <  theta (electron)  <   2.30000D+00  rad
  9.41593D-01  rad  <  theta (positron)  <   2.34159D+00  rad
  Minimal allowed angle between photon and final lepton  =  1.00000D-02 rad.
  Minimal allowed angle between photon and initial lepton=  1.00000D-01 rad.
  acollinearity <   3.14159D+00 rad.
  acoplanarity  <   3.14159D+00 rad.

  Structure of the event: 
  Electron energy (GeV);  theta(electron) (rad.); phi(electron) (rad.) 
  Positron energy (GeV);  theta(positron) (rad.); phi(positron) (rad.) 
  Photon energy  (GeV) ;  theta(photon)  (rad.) ; phi(photon)   (rad.) 

  3.83488D+01   1.33469D+00   1.53144D+00
  4.26590D+01   2.13715D+00   4.65588D+00
  1.39922D+01   1.02513D-01   5.11897D+00
  4.35498D+01   1.14156D+00   8.91211D-01
  4.54257D+01   2.12109D+00   3.98290D+00
  6.02442D+00   3.63628D-01   5.15577D+00
 }}}
\bye